\documentstyle[twoside,fleqn,espcrc2,epsfig]{article}

\newcommand{\AmS}{{\protect\the\textfont2
  A\kern-.1667em\lower.5ex\hbox{M}\kern-.125emS}}

\newcommand{\be}{\begin{equation}}
\newcommand{\ee}{\end{equation}}
\newcommand{\ba}{\begin{eqnarray}}
\newcommand{\ea}{\end{eqnarray}}

\newcommand{\cao}{{\cal O}}
\newcommand{\tao}{\tilde{{\cal O}}}

\title{ 
\vspace{-20.0mm}
\begin{flushright}
\small
HUB-EP-97/62\\JINR-E2-97-280
\end{flushright}
        Gluon propagator and zero-momentum modes on the lattice 
        \thanks{Contribution to Lattice '97, International Symposium, 
                Edinburgh, UK, 1997.
                This research was supported in part under DFG grants 
                Ke 250/13-1 and Mu 932/1-4.} }

\author{G. Damm \address{Fachbereich Physik, Universit\"at Marburg,
		       D-35032 Marburg, Germany },
W.~Kerler $^{\mbox{\scriptsize a,}}$\address{Institut f\"ur Physik, 
Humboldt-Universit\"at, D-10115 Berlin, Germany},
V.K.~Mitrjushkin \address{Joint Institute for Nuclear Research, Dubna, Russia}}

\begin{document}

\thispagestyle{empty}

\begin{abstract}
We investigate the propagators of 4D SU(2) gauge theory in Landau gauge 
by Monte Carlo simulations. To be able to compare with perturbative 
calculations we use large $\beta$ values. There the breaking of the Z(2) 
symmetry causes large effects for all four lattice directions and doing the 
analysis in the appropriate state gets important. We find
that the gluon propagator in the weak-coupling limit is strongly affected 
by zero-momentum modes. 

\end{abstract}

\maketitle

\section{INTRODUCTION} \setcounter{equation}{0}
Starting with Ref.~\cite{mo87} there has been a number of
nonperturbative lattice studies of the gluon propagator in Landau gauge.
However, the impact of zero-momentum modes on the propagators has not been 
analyzed.
Recently one of us has shown that zero-momentum modes may strongly
affect gauge-dependent correlators \cite{m96}.  Therefore, we have
performed simulations in 4D SU2 lattice gauge theory to clarify this
issue.

To be able to compare quantitatively with perturbative calculations large 
$\beta$ values must be used. There the propagators, which gauge fixing 
effectively makes very nonlocal objects, become sensitive to the broken Z(2) 
symmetry states of the deconfinement region. For comparison with perturbative 
results the appropriate one of these states has to be selected.

\begin{figure}[ht]
\vspace*{-2mm}
\epsfig{file=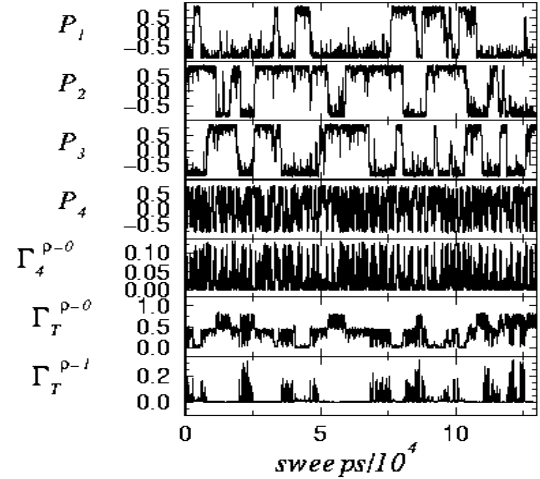,width=7.5cm,height=6.8cm}
\vspace*{-10mm}
\caption{Time history of Polyakov loops and ($\tau=0$) propagators 
	 at $\beta=10$ on $4^3\times 8$ lattice.}
\vspace*{-6mm}

\end{figure}

We use the Wilson action, periodic boundary conditions, fields
$\cao_{\mu}(x)=\frac{1}{2i}(U_{\mu x}-U^{\dag}_{\mu x})$ and the Landau gauge.
The propagators considered are
\be
\Gamma_{\mu}(\vec{p},\tau)= \frac{1}{L_4} \sum_t \mbox{Tr} 
\langle\tao_{\mu}(\vec{p},t+\tau) \tao_{\mu}(-\vec{p},t) \rangle
\label{prop}
\ee
where 
$\tao_{\mu}(\vec{p},\tau)=\frac{1}{V_3}\sum_{\vec{x}}e^{i\vec{p}\cdot\vec{x}}
			 \cao(\vec{x},\tau) $.
Choosing $\vec{p}=(0,0,p_3)$ where $p_3=\frac{2\pi}{L_3}\rho$
the transverse propagator is defined by
$\Gamma_T=\frac{1}{2} (\Gamma_1+\Gamma_2)$. 

\begin{figure}[ht]
\hspace*{-3mm}
\epsfig{file=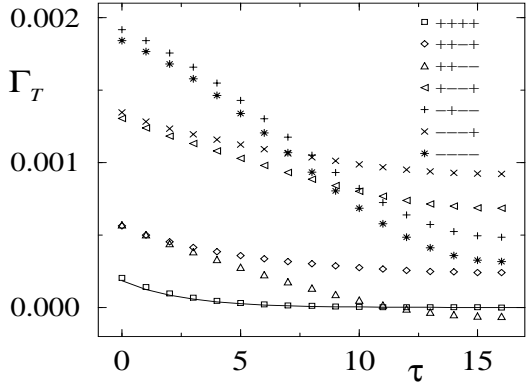,width=11.0cm,height=14.0cm}
\vspace*{-104mm}
\caption{$\Gamma_{T}(\vec{p},\tau)$ in various states for 
 $\rho=1$, $\beta=10$ and $16^3\times 32$ lattice 
 (solid line is (\ref{pprop})).}
\vspace*{-5mm}
\end{figure}

\section{BROKEN Z(2) SYMMETRY} \setcounter{equation}{0}

In our simulations on lattices of sizes $4^3\times 8$, $8^3\times 16$ and 
$16^3\times 32$ we have used Polyakov loops 
$
P_{\mu}=\frac{L_{\mu}}{V_4}\sum_{x\ne x_{\mu}} \mbox{Tr}\prod_{x_{\mu}}U_{\mu x}
$
to monitor the breaking of the Z(2) symmetry. According to the two 
possibilities for each direction (i.e., $P_{\mu}>0$ or $P_{\mu}<0$)
there are $16$ states : $(++++)$, $(+++-)$, 
$\ldots~$, $(----)$. 

The time histories of the $P_{\mu}$, $\Gamma_T$ and $\Gamma_4$ in Fig.~1 
illustrate that the $\Gamma_{\mu}$ are, in fact, strongly affected by
the indicated states.  This phenomenon has been observed on all lattices
considered.  From the numerical analysis we find in more detail that the
observables take different values in different states.  As one can see
in Fig.~2, $(++++)$ gives results consistent with perturbation theory
while otherwise large deviations occur.

\section{FORMS OF PROPAGATORS} \setcounter{equation}{0}

In the limit of large $\beta=4/g^2$ one obtains in lowest--order
approximation
\be
\Gamma_{T}(\vec{p},\tau)\rightarrow\frac{3g^2}{2V_4}\sum_{p_4}
\frac{e^{-ip_4\tau}} {4\sin^2\frac{p_3}{2}+4\sin^2\frac{p_4}{2}}
\label{pprop}
\ee 
provided that $p_3\ne 0$. To handle $\vec{p}=0$ we split off the 
zero-four-momentum part
$ C_{\mu}=\frac{1}{V_4}\sum_{x}\cao_{\mu}(x) $
so that the fields decompose as
$\tao_{\mu}(\vec{0},\tau)= C_{\mu} + \delta\tao_{\mu}(\tau)$
and (\ref{prop}) becomes
\ba
\Gamma_{\mu}(\vec{0}, \tau)&=&\mbox{Tr} \langle C_{\mu}^2\rangle+R_{\mu}(\tau)~;
\label{dprop}
\\
\nonumber \\
R_{\mu}(\tau)&=& \frac{1}{L_4}
\sum_t \mbox{Tr} \langle\delta\tao_{\mu}(t+\tau)
\delta\tao_{\mu}(t)\rangle ~.
\label{rprop}
\ea

\begin{figure}[ht]
\vspace*{-90mm}
\hspace*{-4mm}
\epsfig{file=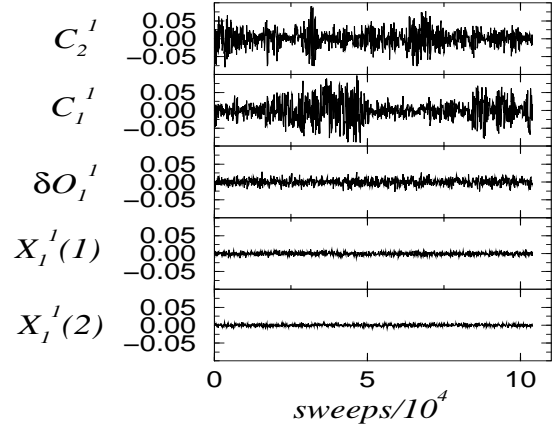,width=14cm,height=15cm}
\vspace*{-19mm}
\caption{Time histories of $C_{\mu}^a$, $\delta \tao_{\mu}^a$, 
 $X_{\mu}^a(1)$ and $X_{\mu}^a(2)$ 
 for $\beta=10$ and $16^3\times 32$ lattice.}
\vspace*{-8mm}
\end{figure}


\noindent The evaluation of (\ref{rprop}) by collective-coordinate
methods only works in the lowest--order approximation and leads to

\be
\Gamma_{\mu}(\vec{0},\tau)\rightarrow \mbox{Tr}\langle C_{\mu}^2\rangle+
\frac{3g^2}{2V_4}\sum_{p_4\ne 0}
\frac{e^{-ip_4\tau}} {4\sin^2\frac{p_4}{2}} ~.
\label{wprop}
\ee

\section{ZERO-MOMENTUM MODES} \setcounter{equation}{0}

Now we restrict the considerations to data of the $(++++)$--state. 
First we note important properties of $C_{\mu}$ by comparing the 
time histories of $C_{\mu}$, $\delta\tao_{\mu}(\tau)$ and $X_{\mu}(\rho)=
\mbox{Re }\frac{1}{V_4}\sum_{x}e^{i\vec{p}\cdot\vec{x}}\cao_{\mu}(x)$
with $\vec{p}=(0,0,\frac{2\pi}{L_3}\rho$) and $\rho\ne 0$, i.e.~of an 
example of a nonzero-momentum part of the fields. Fig.~3 shows
behaviors of components $C_{\mu}^a$, $\delta \tao_{\mu}^a$ 
and $X_{\mu}^a$ (where $C_{\mu} =\sum_a (\sigma^a/2)C_{\mu}^a$ etc.). For
$\delta\tao_{\mu}^a$, $X_{\mu}^a(1)$ and $X_{\mu}^a(2)$ uniform Monte Carlo 
noise is seen. Its magnitude is larger if the three-momentum involved in
the particular quantity gets smaller. For $C_{\mu}^a$ in addition to such 
noise, with magnitude comparable to that of $\delta \tao_{\mu}^a$, 
surprisingly large variations are observed. These variations exhibit
different patterns for different $\mu$, while for different $a$ we find 
essentially the same pattern.

\begin{figure}[ht]
\vspace*{-82mm}
\hspace*{-2mm}
\epsfig{file=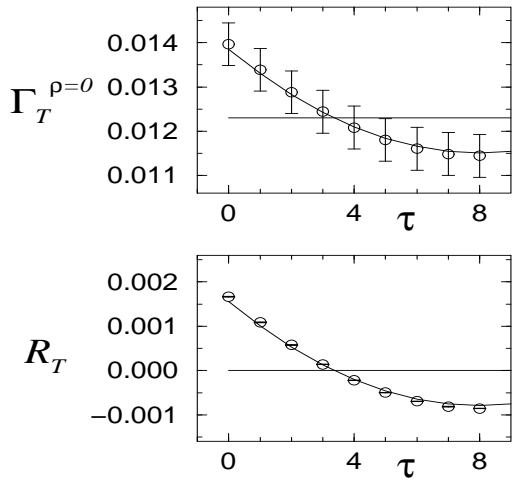,width=13cm,height=15.6cm}
\vspace*{-23mm}
\caption{ $\Gamma_T(\vec{0},\tau)$ and $R_T(\tau)$ compared with 
	 (\ref{wprop}) (curves) and $\mbox{Tr} \langle C^2_{\mu}\rangle$
       (constant line in upper Figure)
	 for $\beta=10$ and $8^3\times 16$ lattice.
}
\vspace*{-7mm}
\end{figure}

In the decomposition of the fields $\tao_{\mu}$ the parts 
$C_{\mu}$ and $\delta \tao_{\mu}$ obviously behave quite differently.
The observed large variations of $C_{\mu}^a$ appear to be a characteristic 
consequence of zero-momentum modes. 

In addition to $\Gamma_T(\vec{0},\tau)$ and $R_T(\tau)$ also the
quantity $\mbox{Tr}\langle C_T^2\rangle$ 
needed for the comparison with (\ref{wprop}) has been 
determined in the simulations. From Fig.~4 it is obvious that our numerical 
results for the transverse propagator with $\vec{p}=\vec{0}$ agree reasonably 
well with that of our lowest-order calculation (\ref{wprop}). It is seen that 
the zero-momentum part $\mbox{Tr}\langle C_T^2\rangle$ 
 is large as compared to the rest, which is related to 
the large variations of $C_{\mu}$ mentioned above. It is also apparent that 
while $\Gamma_T(\vec{0}, \tau)$ exhibits unusually large errors, which
stem from the large variations of $C_{\mu}$, for $R_T(\tau)$ one gets 
errors of usual size. Analogous observations as for the $8^3\times 16$ 
lattice with $\beta=10$ have been made on $4^3\times 8$ with $\beta=10$ and
on $16^3\times 32$ with $\beta=10$ and $\beta=99$. Thus our numerical results
confirm the description we have given.

The importance of zero-momentum modes is quantified by the fact that
$\mbox{Tr}\langle C_{\mu}^2\rangle$ is large as compared to $R_{\mu}(0)$.
To compare different lattice sizes and different values
of $\beta$ the respective values are to be multiplied by $V_3/g^2$ (which 
obviously cancels the extra factor implicit in our definitions). It turns
out that the values of 
$\zeta=(V_3/g^2) \mbox{Tr} \langle C_T^2\rangle$ are much larger than
$\gamma = (V_3/g^2) (R_T(0) - R_T(L_4/2))$.
 From Table 1 it is seen that for increasing $\beta$ this feature 
gets even more pronounced. The same holds for increasing lattice size. Not
only $\zeta$ gets larger but there is also an increase of the ratios
$\zeta/\gamma$.
\begin{table}[h]
\caption{}
\vspace*{-3mm}
\begin{tabular*}{75mm}{@{}l@{\extracolsep{\fill}}llll}
\hline
 lattice      & $\beta$ & $\zeta$ &  $\gamma$ \\
\hline

$4^3\times 8$ &  10  &  6.30 (16)  &  1.610 (5)   \\
$8^3\times 16$ &  10 &  15.7 (7) &  3.229 (17)   \\
$16^3\times 32$ &  10 &  57 (10) &  6.03 (22)     \\
$16^3\times 32$ &  99 & 375 (86) &  4.98 (19)     \\
\hline
\end{tabular*}
\vspace*{-6mm}
\end{table}

To show that the usual determination of effective masses $m (\tau)$ can 
be misleading if zero-momentum modes are present we have also calculated
$m (\tau )$ from
\be
\frac{\cosh (m (\tau )(\tau +1-\frac{L_4}{2}))}
{\cosh (m (\tau )(\tau -\frac{L_4}{2}))}
= \frac{\Gamma_T(\vec{0},\tau +1)}{\Gamma_T(\vec{0},\tau )}~.
              \label{m_eff}
\ee
 From Fig.~5 it is seen that even at very weak coupling, where the situation 
clearly is described by (\ref{wprop}), a gluon mass is imitated. Its
decrease with lattice size actually reflects the increase of the 
zero-momentum contribution. Because $m(\tau)L_{\mu}$ shows 
little dependence on lattice size, the present results may also be considered
from the point of view of finite temperatures where screening masses 
are determined \cite{hkr95}. 

To summarize, at $\beta$--values considered here such masses are fake.
The question of the role of the zero--momentum modes at smaller values
of $\beta$ (in the physical region) needs further study.

\begin{figure}[t]
\vspace*{2mm}
\hspace*{-8mm}
\epsfig{file=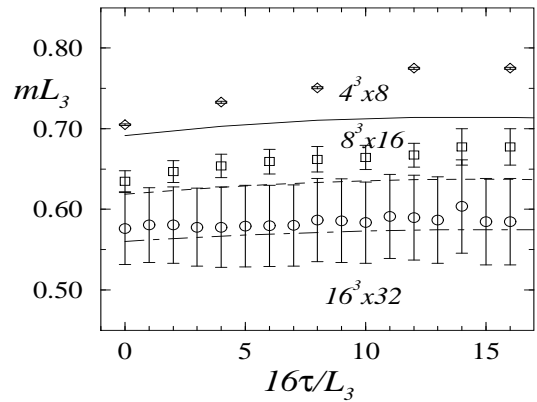,width=13.0cm,height=16.0cm}
\vspace*{-124mm}
\caption{Misleading results from determinations of effective masses
        from eq.~(\ref{m_eff}) for $\vec{p}=\vec{0}$, $\beta=10$.
	  Curves are based on eq.~(\ref{wprop}).
}
\vspace*{-5mm}
\end{figure}

\vspace*{3mm}
One of us (W.K.) wishes to thank M.~M\"uller-Preussker and his group for 
their kind hospitality.

\end{document}